\documentclass{aa} 

\usepackage{natbib}
\usepackage{graphicx}
\usepackage{txfonts}
\newcommand{\Mpc}{$h^{-1}$\thinspace Mpc}

\newcommand{\vmh}{h^{-1}\mathrm{Mpc} }

\begin{document}

\title{The evolution of high-density cores of the BOSS Great Wall superclusters}

\author{
Maret Einasto \inst{1} \and
Peeter Tenjes \inst{1}  \and
Mirt Gramann \inst{1} \and
Heidi Lietzen \inst{1,2} \and
Rain Kipper \inst{1} \and
Lauri Juhan Liivamägi \inst{1} \and
Elmo Tempel \inst{1,3} \and
Shishir Sankhyayan \inst{1} \and
Jaan Einasto \inst{1,3,4}
}

\institute{  
Tartu Observatory, Tartu University, Observatooriumi 1, 61602 T\~oravere, Estonia
\and
The National Library of Finland, Unioninkatu 36, University of
Helsinki, 00014 Helsinki, Finland 
\and
Estonian Academy of Sciences, Kohtu 6, 10130 Tallinn, Estonia
\and
ICRANet, Piazza della Repubblica 10, 65122 Pescara, Italy
    }

\authorrunning{Einasto, M. et al. }

\offprints{Einasto, M.}

\date{ Received   / Accepted   }

\titlerunning{BGW collapse}

\abstract
{
High-density cores (HDCs) of  galaxy superclusters that embed rich clusters and groups of galaxies
are the earliest large objects to form in the cosmic web, 
and the largest objects that may collapse
in the present or future. 
}
{
We  aim to study the dynamical state and possible  evolution
of the HDCs in the BOSS Great Wall (BGW) superclusters at redshift  $z \approx 0.5$
from the CMASS (constant mass) galaxy sample,
based on the Baryon Oscillation Spectroscopic Survey (BOSS) in order  to understand the growth and evolution of structures in the Universe.
}
{
We analysed the luminosity density distribution in the BGW superclusters to determine 
the  HDCs in them. 
We derived the density contrast values for the spherical collapse model 
in a wide range of redshifts and used these values 
to study the dynamical state and possible evolution of the HDCs 
of the BGW superclusters.  The masses of the HDCs were calculated 
using stellar masses of galaxies in them.
We found the masses and radii of the turnaround and future collapse
regions in the HDCs of the BGW superclusters and compared 
them with those of local superclusters.
}
{
We determined eight HDCs in the BGW superclusters.
The masses of their turnaround regions 
are in the range of $M_{\mathrm{T}} \approx 0.4 - 3.3\times~10^{15}h^{-1}M_\odot,$
and radii are in the range of  $R_{\mathrm{T}} \approx 3.5 - 7$~\Mpc. 
The radii of their future collapse regions
are in the range of  $R_{\mathrm{FC}} \approx 4 - 8$~\Mpc. Distances between individual
cores in superclusters are much larger: of the order of $25 - 35$~\Mpc.
The richness and sizes of the HDCs are comparable with those
of the HDCs of the richest superclusters in the local Universe.
}
{
The BGW superclusters will probably evolve to several poorer
superclusters with masses similar to those of the local superclusters. 
This may weaken the tension with the
$\Lambda$CDM model, which does not predict a large number of very rich and
large superclusters
in our local cosmic neighbourhood, and explains why there are no 
superclusters as elongated as those in the BGW in the local Universe. 
}


\maketitle
\section{Introduction}

In the complex hierarchical network 
of galaxies, galaxy groups, clusters, and superclusters, called 
the cosmic web,
the largest systems are galaxy superclusters and supercluster complexes,
where several very rich superclusters form chains or planes
(as, for example, in the Sloan Great Wall; hereafter SGW) 
\citep{1956VA......2.1584D, 1978MNRAS.185..357J, 
2007A&A...462..397E, 2011ApJ...736...51E, 2015A&A...575L..14C,
2015A&A...580A..69E, 2016A&A...588L...4L, 2016A&A...595A..70E}.
Rich superclusters embed high-density cores (hereafter HDCs) with one or 
several rich galaxy clusters, while poor superclusters typically do not have
HDCs \citep{2007A&A...464..815E}. 
The formation of the cosmic web started from tiny 
density perturbations in the very early Universe
\citep{1978MNRAS.185..357J, 1988Natur.334..129K}. 
The skeleton of the cosmic web is
fixed by processes during or just after the inflation \citep{1988Natur.334..129K}. 
The positions of high-density
peaks and voids do not change much during the cosmic evolution, only the amplitude of
over- and under-densities grows with time
\citep[][and references therein]{1996Natur.380..603B, 2009LNP...665..291V,
2011A&A...531A.149S, 2019A&A...623A..97E, 2021A&A...652A..94E}. 
Superclusters or their HDCs  are the largest systems in the cosmic web 
that may eventually collapse 
\citep{1956VA......2.1584D, 1978MNRAS.185..357J, 
2015A&A...580A..69E, 2015A&A...575L..14C, 2016A&A...588L...4L, 2016A&A...595A..70E,
2018A&A...620A.149E, 2021A&A...649A..51E}.

Observationally, the studies of protoclusters, progenitors of the present-day
galaxy clusters, have shown that they start to form in regions
of the highest density in the cosmic density field and 
are already present at redshifts of $z \approx 6$ 
\citep[][and references therein]{2016ApJ...826..114T, 2016A&ARv..24...14O,
2018MNRAS.474.4612L}. Galaxy protoclusters are the first  sites 
of galaxy and cluster formation in a  high-redshift universe
\citep{2017ApJ...844L..23C, 2018Natur.553...51M, 2018ApJ...861...43P,
2020MNRAS.496.3169A, 2022ApJ...926...37M}.
Typically, the most luminous galaxy clusters 
are located in the HDCs
of rich superclusters  \citep{1978MNRAS.185..357J}.

The advent of deep surveys such as the 2dFRS (2dF Galaxy Redshift Survey), 
SDSS (Sloan Digital Sky Survey), and CMASS (constant mass) 
makes it possible to compile supercluster catalogues in a wide redshift interval
\citep{2007A&A...462..397E, 2012A&A...539A..80L, 2014MNRAS.445.4073C}
or to determine individual superclusters  at high redshifts 
\citep{2007MNRAS.379.1546T, 2011A&A...532A..57S, 2016A&A...592A...6P,
2016ApJ...821L..10K, 2018A&A...619A..49C}.
To identify elements of the cosmic web,
various methods have been used \citep[see e.g. ][for a review]
{2018MNRAS.473.1195L}. Superclusters of galaxies have been defined
using various methods and criteria, briefly described in \citet{2020A&A...641A.172E}.
Superclusters  can be defined as the largest connected, relatively isolated  
objects above a certain density
level in the cosmic web \citep{1994MNRAS.269..301E,2001AJ....122.2222E, 2012A&A...539A..80L}. 
We also use this definition in the present paper.
\citet{2011MNRAS.415..964L} and \citet{2015A&A...575L..14C} defined superclusters
as objects that will eventually collapse in the future. Comparison
of these definitions shows that the latter definition
tends to select HDCs of superclusters obtained using the first definition
\citep{2015A&A...575L..14C, 2015A&A...580A..69E, 2016A&A...595A..70E}.
Rich protoclusters, determined at high redshifts, may mark the cores of proto-superclusters  
\citep{2016A&A...595A..70E,
2018A&A...619A..49C, 2020MNRAS.496.3169A, 2022ApJ...926...37M}.

In the $\Lambda$ cold dark matter ($\Lambda$CDM) Universe
at redshifts of $z \approx 0.5$ and below, the growth of structures slows down under the influence
of the dark energy \citep{2008ARA&A..46..385F, 2021A&A...652A..94E}.  
At this redshift, several superclusters
have been discovered \citep{2011A&A...532A..57S, 2018MNRAS.475.4148G}.
We discuss some of them in Sect.~\ref{sect:disc}. 
Typically, these 
superclusters resemble local medium-rich superclusters or the
HDCs of rich superclusters \citep{2018A&A...620A.149E}. 
However, 
\citet{2016A&A...588L...4L} reported a
discovery of a very massive extended supercluster complex at redshift of
$z = 0.45$ named as the BOSS Great Wall (hereafter BGW) 
using the CMASS sample of the SDSS III (SDSS-III) 
\citep{2011AJ....142...72E, 2013MNRAS.435.2764M, 2016MNRAS.455.1553R}. 
The BGW consists of four superclusters, two of them are extraordinarily 
rich and more elongated than any local supercluster
\citep[Fig.~\ref{fig:radec} and ][]{2017A&A...603A...5E}. 
\citet{2017A&A...603A...5E} analysed
the morphology, luminosity, and mass of the BGW superclusters.
They found that 
these superclusters form a complex that is the richest supercluster
complex known so far,  being richer than the richest supercluster
complex in the local Universe, the SGW \citep{2016A&A...595A..70E}.

Extreme objects such as the BGW 
usually provide constraints for theories.
It is not clear whether systems with sizes, richness, and morphologies similar to
those of the BGW can be reproduced in the $\Lambda$CDM model
or agree with the Gaussian initial conditions 
\citep{2007A&A...476..697E, 2011ApJ...736...51E, 2011MNRAS.417.2938S,
2012ApJ...759L...7P, 2017A&A...603A...5E}. It is also not clear why 
there are no such very rich supercluster complexes with
very  elongated systems in the local Universe.
This makes galaxy superclusters and their HDCs at different redshifts
unique objects with regard to studies of the formation
and evolution of the cosmic web and testing cosmological 
models and the cosmological principle of homogeneity and isotropy
of the Universe.
If  during the future evolution BGW superclusters break into  smaller systems
defined by  the  HDCs of superclusters, similarly to the 
richest local superclusters such as the Shapley or the Corona Borealis
superclusters or the superclusters in the SGW, 
then the tension with models may weaken or disappear 
\citep{2015A&A...575L..14C, 2016A&A...595A..70E, 2021A&A...649A..51E}. 

In this study, our aim is to obtain a better understanding 
of the evolution of the largest structures in the Universe.
We focus on a possible evolution 
of the BGW superclusters and especially their HDCs. 
Could it be possible that  the BGW superclusters fall apart and form smaller 
systems during evolution?  
Answering this question will help to clarify whether the tension with 
the $\Lambda$CDM model will disappear or weaken. 
With this aim in mind, we located the HDCs of the BGW superclusters 
and analysed  the possible 
evolution of their masses and sizes using spherical collapse model.
We determined masses of the HDCs of superclusters 
using stellar masses of high-mass galaxies, as 
described in detail in \citet{2017A&A...603A...5E}.

For this aim we first needed to derive 
the parameters of the spherical collapse model for a wide range
of redshifts and then apply them to search for possible collapsing regions in the HDCs
of the BGW superclusters. We compare the properties of the turnaround and future collapse
regions in the HDCs of the BGW superclusters, and  in superclusters at different redshifts.
Superclusters as a whole are very elongated, but their HDCs are 
close to  spherical, and therefore we can use the spherical collapse model
to analyse their dynamical state and possible evolution \citep{2021A&A...649A..51E}.
We assumed  the standard (WMAP) cosmological parameters: the Hubble parameter $H_0=100~ 
h$ km~s$^{-1}$ Mpc$^{-1}$, the matter density $\Omega_{\rm m} = 0.27$, and the 
dark energy density $\Omega_{\Lambda} = 0.73$ \citep{2011ApJS..192...18K}.
    
\section{Data on the BGW superclusters}
\label{sect:data} 

\begin{figure}%[ht]
\centering
\resizebox{0.44\textwidth}{!}{\includegraphics[angle=0]{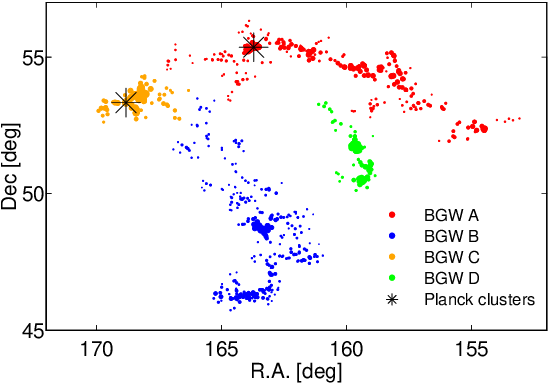}}
\caption{
Distribution of galaxies in BGW superclusters in the sky plane. 
Red dots denote galaxies in BGW A, blue dots in BGW B,
yellow dots in BGW C, and green dots in BGW D.
Symbol sizes are proportional to the value of the density field at the
location of a galaxy. Stars denote the location of Planck clusters.
}
\label{fig:radec}
\end{figure}

\begin{table}
\caption{Data on superclusters in BGW.} 
\centering
\begin{tabular}{crrrrr}
\hline\hline
(1)&(2)&(3)&(4)&(5)& (6) \\      
\hline 
ID &  $N_{\mathrm{gal}}$ &  Extent & $M$ & $M/L$ & $D8_{\mathrm{mean}}$ \\
 &  & $h^{-1}$Mpc &  $10^{15}h^{-1}M_\odot$ & $h\,M_\odot$/$L_\odot$  & 
 $\langle \rho_\mathrm{L} \rangle$  \\
\hline       
A & 255 & 186.1 & 21  &277 & 9.1  \\
B & 303 & 172.9 & 11  &182 & 9.3  \\
C & 73  &  63.8 &  7  &407 &10.2  \\
D & 71  &  90.6 &  3  &118 & 9.3  \\
\hline             
\end{tabular}
\tablefoot{
Columns are as follows:
(1) notation; 
(2) number of galaxies in a supercluster from CMASS catalogue, $N_{\mathrm{gal}}$;
(3) maximal extent of a supercluster;
(4) mass of a supercluster;
(5) mass-to-light ratio of a supercluster;
(6) mean luminosity density in superclusters, $D8_{\mathrm{mean}}$,
in units of the mean luminosity density calculated
with an $8$~\Mpc\ smoothing scale (see text).
} 
\label{tab:bgwscl}
\end{table}    

To determine galaxy superclusters at redshifts $z \approx 0.5,$ 
\citet{2016A&A...588L...4L} used data from the CMASS galaxy sample
based on the Baryon Oscillation Spectroscopic Survey 
\citep[BOSS;][]{2011ApJS..193...29A, 
2011AJ....142...72E, 2012AJ....144..144B, 2013AJ....145...10D}. 
The CMASS sample includes data about massive and luminous galaxies 
in the redshift range $0.43<z<0.7$. Their stellar mass is 
approximately constant up to $z\sim 0.6$ \citep{2013MNRAS.435.2764M}. 
These are galaxies from the 
high-mass end ($M>10^{11}~M_{\odot}$) of the red-sequence galaxies,
which do not evolve over the CMASS redshift range.
For details of our sample, we refer the reader to \citet{2016A&A...588L...4L}.

Galaxy superclusters were determined using the
luminosity-density field 
following the same procedure that was used in \citet{2012A&A...539A..80L}. 
The luminosities of galaxies were weighted to set the mean luminosity density the 
same through the whole distance range. 
Then, the luminosity density field was calculated in a 3\,$h^{-1}$Mpc grid 
with an 8\,$h^{-1}$Mpc smoothing scale
using a $B_3$ spline kernel:
\begin{equation}
    B_3(x) = \frac{1}{12} \left(|x-2|^3 - 4|x-1|^3 + 6|x|^3 - 4|x+1|^3 + |x+2|^3\right).
\end{equation}
The calculation of the luminosity-density field is described in detail in
\citet{2012A&A...539A..80L}.

The BGW superclusters were extracted as connected volumes above the density
level $D > 6$ times the mean luminosity density of the CMASS sample,
$\ell_{\mathrm{mean}}$ = 
5$\cdot10^{-4}$ $\frac{10^{10} h^{-2} L_\odot}{(\vmh)^3}$) 
\citep{2016A&A...588L...4L}.
The BGW supercluster complex 
consists of two very rich superclusters with  maximal extent 
(the maximal distance between galaxies in the supercluster) 186\,$h^{-1}$Mpc 
(supercluster A) and 
173\,$h^{-1}$Mpc (supercluster B), 
and two moderately large 
superclusters (superclusters C and D) 
with  maximal extent 64 and 91\,$h^{-1}$Mpc. 
Data concerning the BGW superclusters are given in Table~\ref{tab:bgwscl}.
In Table~\ref{tab:bgwscl}, the masses are as those derived in \citet{2017A&A...603A...5E} 
and  explained in Sect.~\ref{sect:masses}. 
Figure~\ref{fig:radec} presents the sky distribution of galaxies 
in the BGW superclusters.         

\citet{2017A&A...603A...5E} identified two Planck SZ clusters 
in the BGW region using data from
the Second Planck Catalogue of Sunyaev-Zeldovich Sources 
\citep{2016A&A...594A..27P}: 
PSZ2~731 in the  BGW C and PSZ2~735 in the supercluster BGW A.
Their location is shown in Fig.~\ref{fig:radec}.

\section{Methods}

\subsection{Spherical collapse model} 
\label{sect:sph} 

\begin{figure*}[ht]
\centering
\resizebox{0.44\textwidth}{!}{\includegraphics[angle=0]{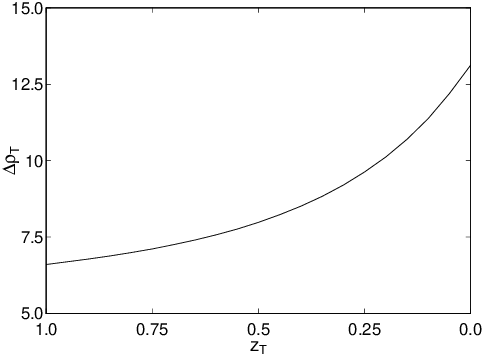}}
\resizebox{0.44\textwidth}{!}{\includegraphics[angle=0]{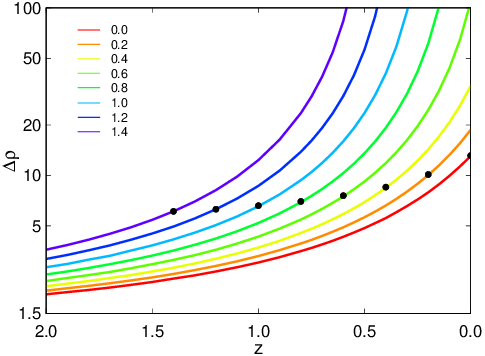}}
\caption{
Turnaround density contrast $\Delta\rho_{\mathrm{T}}$ versus redshift 
at which  turnaround occurs, $z_T$ (left panel) and 
evolution 
of the density contrast $\Delta\rho(t)$ 
in the spheres, where the turnaround happens
at redshifts $z = 0.0 - 1.4$  (right panel). Corresponding redshifts are shown in the
figure.
Points on each curve show the turnaround density contrast $\Delta\rho_{\mathrm{T}}$
at a corresponding redshift.
}
\label{fig:turn}
\end{figure*}

The spherical collapse model describes the evolution of a spherical 
perturbation in an expanding universe. 
This model was studied, for example,  by
\citet{1980lssu.book.....P},
\citet{1984ApJ...284..439P}, and \citet{1991MNRAS.251..128L}.  
In the standard models with cosmological constant, 
the formation of structures in the cosmic web slows down at redshifts  $z \approx 0.5$
due to the influence of the dark energy \citep{2008ARA&A..46..385F}. 
At the present epoch, the largest bound structures are just forming.  
During the future evolution
these bound systems separate from each other at an accelerating rate, 
forming isolated 'island universes'
\citep{2002PhRvD..65l3518C, 2006MNRAS.366..803D, 2009MNRAS.399...97A}.

The dynamics of a spherical  shell is determined 
by the mass in its interior. 
For a spherical volume, 
the ratio of the density to the mean density (overdensity)
$\Delta\rho = \rho/\rho_{\mathrm{m}}$ at different redshifts can be calculated as
\begin{equation}
\Delta\rho=6.88\,\Omega_\mathrm{m0}^{-1}\left(\frac{M}{10^{15}h^{-1}M_\odot}\right)
        \left(\frac{R}{5h^{-1}\mathrm{Mpc}}\right)^{-3}\cdot (1+z)^{-3},
\label{eq:sph}
\end{equation}
 where $\Omega_{m0}$ is the matter density parameter at present.
From Eq.~(\ref{eq:sph}), we can find the mass of a structure as
\begin{equation}
M(R)=1.45\cdot10^{14}\,\Omega_\mathrm{m0}\Delta\rho\left(R/5h^{-1}\mathrm{Mpc}\right)^3 \cdot (1+z)^3 h^{-1}M_\odot.
\label{eq:mass1}
\end{equation}

The evolution of a spherical shell
has several essential epochs, each of which has a characteristic density contrast
\citep{2015A&A...575L..14C, 2015A&A...581A.135G}.
These density contrasts can be used to derive the relations 
between the radius of a perturbation and its interior mass for each essential epoch.
In this study, we focused on the turnaround and future collapse.
Turnaround is defined as an epoch in the evolution of a spherical perturbation 
when the sphere stops expanding 
together with the universe and the collapse begins. 
Turnaround has a characteristic density contrast which changes with redshift.
This means that objects for which the density contrast at a given redshift is not sufficient for
the turnaround at this redshift may eventually reach turnaround and
experience collapse in the future if their density
contrast is high enough (the so-called future collapse).

Next, we briefly describe the evolution of the density contrast. We start with turnaround
epoch.
The evolution of a density perturbation is affected by the density in it and
the spherically averaged radial velocity around it. 
The spherically averaged radial velocity around a system 
in the shell of radius $R$ can be written as $u = HR - v_{\mathrm{pec}}$, 
where $v_{\mathrm{H}} = HR$ is the Hubble expansion velocity and 
$v_{\mathrm{pec}}$ is 
the averaged radial peculiar velocity towards the centre of the system. 
At the turnaround, the peculiar velocity $v_{\mathrm{pec}} = HR$ and $u = 0$. 
The peculiar velocity $v_{\mathrm{pec}}$ 
is directly related to the overdensity $\Delta\rho$.

As systems evolve, their density increases, and in the standard models with 
cosmological constant, the density contrast at the turnaround, $\Delta\rho_T$,  also
increases during the evolution. We show this
 in Fig.~\ref{fig:turn} 
for $\Omega_{\mathrm{m0}} = 0.27$ and $\Omega_{\mathrm{\Lambda}} = 0.73$, which we used in the present study. 
Figure~\ref{fig:turn} (left panel) shows how the turnaround density contrast
$\Delta\rho_{\mathrm{T}}$ increases with redshift.
The overdensity at the turnaround at redshift $z = 0.0$
is  $\Delta\rho_{\mathrm{T}} =  13.1$. 
The overdensity at the turnaround at the redshift $z = 0.5$
(the redshift of the BGW superclusters),
$\Delta\rho_T = 8.0$,  and at the redshift $z = 1.0$ $\Delta\rho_T = 6.6$.

The value of the overdensity at the turnaround at a given redshift
can be used to calculate the mass of a structure 
at the turnaround at that redshift.
For example,  at the redshift $z = 0.5,$
the mass of a structure at the turnaround  is
\begin{equation}
$$
M_\mathrm{T}(R)=1.01\cdot10^{15}\left(R/5h^{-1}\mathrm{Mpc}\right)^3h^{-1}M_\odot.
%M_T(R) =  3.1 \times 10^{14} R_{5}^3 \,\, h^{-1} M_\odot.
$$
\label{eq:z05}
\end{equation}

In addition, in Fig.~\ref{fig:turn} (right panel)
we show the evolution 
of the density contrast  $\Delta\rho(t)$ for cases where the turnaround occurs
at a range of redshifts from $z = 0.0$ to $z= 1.4$.
From this figure, one can see, for example, that if the turnaround occurred at
redshift $z \approx 0.4$ then the present density contrast for such a structure is
$\Delta\rho \approx 30$ \citep[see also][in which this characteristic density contrast
around rich clusters in the A2142 and the Corona Borealis superclusters was found]
{2020A&A...641A.172E, 2021A&A...649A..51E}.

Another characteristic epoch in the evolution of the density contrast
is the future collapse. 
The superclusters that have not reached the turnaround at a given 
redshift may eventually turn around and collapse in the future 
if their density contrast 
is high enough \citep{2006MNRAS.366..803D, 2015A&A...575L..14C}. 
These studies showed that the minimum density at redshift $z = 0.0$ 
for a shell to remain bound is 
$\Delta\rho_{\mathrm{FC}} =  8.73$. This corresponds to the turnaround at 
infinite distant future. 
At the redshift $z = 0.5,$ 
the overdensity for the collapse at 
infinite distant future is $\Delta\rho_{\mathrm{FC}} =  4.02$, 
and at the redshift $z = 1.0$, 
the overdensity is
$\Delta\rho_{\mathrm{FC}} = 2.72$.

\begin{figure*}[ht]
\centering
\resizebox{0.44\textwidth}{!}{\includegraphics[angle=0]{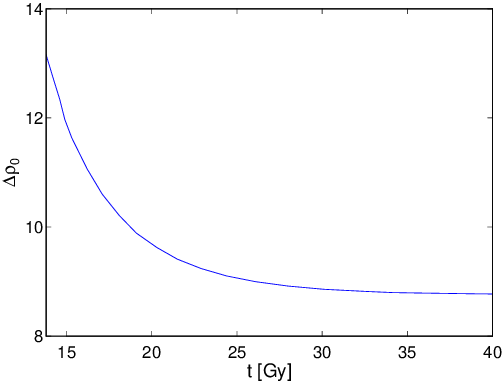}}
\resizebox{0.44\textwidth}{!}{\includegraphics[angle=0]{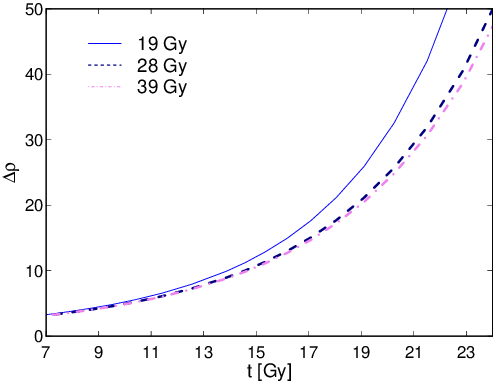}}
\caption{
Density contrast $\Delta\rho_{\mathrm{0}}$ at present versus time $t$
at which  turnaround occurs (left panel) and  
evolution of the density contrast $\Delta\rho$
as a function of time $t$ (the age of the Universe)
for three different future collapse times
($19$, $28$, and $39$~Gys, right panel).
}
\label{fig:fc}
\end{figure*}

For intermediate density contrasts $13.1 > \Delta\rho > 8.73,$ 
future collapse (turnaround and contraction after that) starts at a certain 
time in the future. 
The density contrasts for the collapse at 
certain times in the future can be calculated by integrating equations 
for the evolution of a spherical density contrast.
We refer the reader to \citet{1998ApJ...508..483W} and \citet{2010JCAP...10..028L}
for details.
We show the density contrast at present 
as a function of future collapse start time 
(turnaround time in the future) in the left panel of Fig.~\ref{fig:fc}. 
The right panel of Fig.~\ref{fig:fc} shows the evolution of the density contrast 
as a function of time (the age of the Universe) 
for three different future collapse times, as given in the figure. 
The starting time in this figure is approximately 
$5$~Gyrs past from the present time (this corresponds approximately to the redshift $z = 0.5$,
which is the redshift of the BGW).
Figure~\ref{fig:fc}
shows that at $z = 0.5$ ($t \approx 8.8$~Gyrs) 
density contrasts are $4.4$, $4.09$, and $4.06$,
if the collapse occurs in $19$, $28$, or $39$~Gyr.  In other words, the density contrast
only varies a little.

In our study we used the density contrast for the collapse at 
infinite distant future, $\Delta\rho_{\mathrm{FC}} =  4.02$, 
to calculate the minimum mass of the structure at the redshift $z = 0.5,$ which will 
turn around and collapse in the future as
\begin{equation}
M_\mathrm{FC}(R)=5.29\cdot10^{14}\left(R/5h^{-1}\mathrm{Mpc}\right)^3h^{-1}M_\odot.
\label{eq:mrfvs}
\end{equation}
We used this relation to find the masses and radii of the regions 
in the HDCs of the BGW superclusters that may eventually collapse during their
evolution.

\subsection{Determination of  masses of HDCs in the BGW superclusters}
\label{sect:masses}

In the BGW superclusters, only the brightest galaxies can be observed.
To determine the mass of the BGW superclusters and their HDCs,
we apply the  relation between the stellar masses of the
first ranked galaxies in haloes,  $M_*$, and 
the virial masses of the haloes to which these galaxies belong, $M_{\mathrm{halo}}$
\citep{2010ApJ...710..903M}:
\begin{equation}\label{eq:mass}
\frac{M_{*}}{M_{\mathrm{halo}}}=2\left(\frac{M_{*}}{M_{\mathrm{halo}}}
\right)_0 \left[\left(\frac{M_{\mathrm{halo}}}{M_1}
\right)^{-\beta}+\left(\frac{M_{\mathrm{halo}}}{M_1}\right)^\gamma\right]^{-1},
\end{equation}
where $(M_*/M_{\mathrm{halo}})_0 = 0.0254$ is the normalisation of the stellar-to-halo mass 
relation, $M_1=10^{11.95}$ is a characteristic mass, 
and $\beta=1.37$ and $\gamma=0.55$ are the slopes of the low 
and high mass ends of the relation, respectively \citep[][Table 6]{2010ApJ...710..903M}. 
The sum of the halo masses gives us an estimate of the lower limit
of the supercluster mass. 
This is the same 
procedure as was applied in \citet{2016A&A...588L...4L}
and \citet{2017A&A...603A...5E} to find the masses of the BGW superclusters.

The stellar masses of BOSS galaxies were obtained from the Portsmouth galaxy product 
\citep{2013MNRAS.435.2764M}, which is based on the stellar population 
models by \citet{2005MNRAS.362..799M} and \citet{2009MNRAS.394L.107M}. The Portsmouth 
product uses an adaptation of the publicly available Hyper-Z code 
\citep{2000A&A...363..476B} to perform a best fit to the observed {\it{ugriz}} 
magnitudes of BOSS galaxies, with the spectroscopic redshift determined by the BOSS pipeline. 
The stellar masses used in this work were computed assuming 
the Kroupa initial mass function. 

In calculations of the masses of superclusters, we
only used galaxies with stellar masses of $\log(M_*/M_\odot) \geq 11.3$,
which is the  completeness limit of the CMASS sample \citep{2013MNRAS.435.2764M}. 
We assumed that galaxies in the CMASS sample with $\log(M_*/M_\odot)>11.3$
are the  central  galaxies of haloes.
This is based on comparison with 
the SDSS main sample of galaxies as follows. 
We used the magnitude-limited friend-of-friends group catalogue from the SDSS DR10
main sample
by \citet{2014A&A...566A...1T} to select galaxies 
in superclusters with the luminosity density $D8 \geq 5$
in the distance bin from 180 to 270\,$h^{-1}$Mpc.  
From this sample, we determined a BOSS CMASS-like high-mass sample of galaxies 
with a stellar mass limit of $\log(M_*/M_\odot) \geq 11.3$
and found that 87\,\% of all galaxies in this high-mass sample
are the most luminous,  central galaxies 
in the friend-of-friends groups, or they are single galaxies 
(the  central galaxies of groups
with satellite galaxies which are too faint to be 
observed in the SDSS). Therefore, we can assume that the high-mass galaxies 
in the BOSS sample are the  central galaxies in groups or rich clusters.  
Comparison with local galaxies suggests that this may introduce an error in mass estimates
of the order of about 10--15\,\%, considering that some massive galaxies may be
members of the same group and not the  central galaxies of different groups.

There are also haloes whose  central galaxies have lower stellar
masses than the  completeness limit $\log(M_*/M_\odot) = 11.3$.
To take into account the mass in these haloes, we applied scaling based on the analysis
of the SDSS main sample of galaxies. 
We again used the data concerning galaxies in superclusters with the luminosity density $D8 \geq 5$
in the distance bin from 180 to 270\,$h^{-1}$Mpc, and found that the  
ratio between the total stellar mass in  high-mass  central galaxies 
and the total stellar mass in all  central galaxies in the SDSS 
superclusters is 0.082. Therefore, we
used this ratio to scale our supercluster mass estimates. 
For details, we refer the reader to \citet{2016A&A...588L...4L} and \citet{2017A&A...603A...5E}.

To obtain the total masses of the BGW superclusters, \citet{2017A&A...603A...5E} 
used the stellar masses of high-mass galaxies and the scaling as applied here 
to correct the mass estimate for incompleteness. They additionally 
used several methods 
to determine masses of the BGW superclusters and of the SGW
in \citet{2016A&A...595A..70E}.
In addition to the stellar masses of the central galaxies, they
applied mass-to-light ratios and combined morphological and physical
parameters of superclusters to determine their masses. 
\citet{2017A&A...603A...5E} showed that various methods gave very similar
masses of superclusters. For example, 
the mass of the BGW A supercluster, obtained with different methods,
was in the range of
$21 - 23\times~10^{15}h^{-1}M_\odot$, and
the mass of the BGW B supercluster in the range of
$11 - 18\times~10^{15}h^{-1}M_\odot$
\citep{2017A&A...603A...5E}. 
In the case of SGW those mass estimates 
that used stellar masses
of galaxies gave similar masses as those methods that used dynamical masses of groups.
With these methods, \citet{2016A&A...595A..70E} showed  that 
the total  mass of SGW is 
in the range of $22 - 24\times~10^{15}h^{-1}M_\odot$.
Therefore, different mass estimates are in good agreement, which suggests that we do
not have strong biases in the mass estimate of the BGW superclusters;
although, we only used data on high-mass  galaxies as this sample is complete.

We assume that the high-mass galaxies represent the central galaxies
of rich clusters. The number of such clusters in the BGW superclusters
is comparable to the number of rich clusters in local rich superclusters
such as the superclusters in the SGW or the Corona Borealis supercluster
\citep{2016A&A...595A..70E, 2021A&A...649A..51E}.
However, as we show in Sect.~\ref{sect:collcores}, with this mass estimate
we cannot determine the mass
of one possible high-density region in the BGW
supercluster that does not contain any high-mass galaxy.
Such regions may be similar to, for example, those in the tail of the
A2142 supercluster that contain several regions with merging groups
\citep{2018A&A...620A.149E, 2020A&A...641A.172E}.

We find the distribution of mass around the possible centres of the 
HDCs of the BGW superclusters (see Sect.~\ref{sect:collapse}).
The comparison of the observed mass distribution with the predictions of
the spherical collapse model gives us the masses and radii of the turnaround 
and future collapse regions. 

\begin{figure*}%[ht]
\centering
\resizebox{0.75\textwidth}{!}{\includegraphics[angle=0]{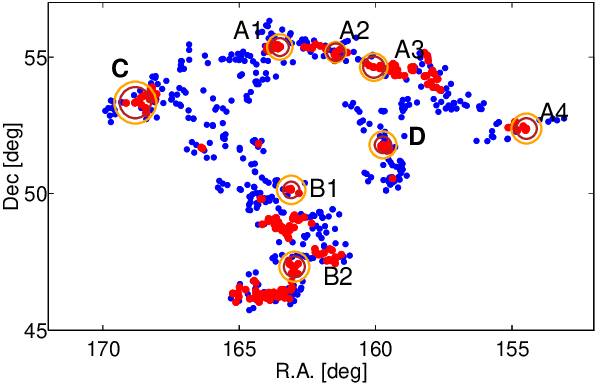}}
\caption{
Distribution of galaxies in BGW superclusters in the sky plane. 
Red dots denote galaxies in the possible HDCs of each supercluster, and 
blue dots show galaxies in outskirts.
Luminosity limits for HDCs are as follows: $D8_{HDC} \geq 15$ in BGW~A,
 $D8_{HDC} \geq 11$ in BGW~B, $D8_{HDC} \geq 14.5 $ in BGW~C, and $D8_{HDC} \geq 16$ in BGW~D,
where $D8$ is the value of the luminosity density (see text).
Dark red circles show the turnaround regions in each HDC, and orange circles show the future collapse 
regions.
Numbers with labels denote HDCs in each BGW supercluster.
}
\label{fig:radeccen}
\end{figure*}

For comparison with local superclusters, we also estimated the mass of the 
HDCs of the BGW superclusters at redshift $z = 0.0$.
In this comparison, we apply several estimates for the masses and radiii of the collapsing
regions. We used the predictions from simulations, which show that 
the mass of massive haloes increases two times
from redshift $z = 0.5$ to the present \citep{2015JKAS...48..213K}.
We applied this estimate to find the masses of HDCs of superclusters at $z = 0.0$
and applied a mass-to-radius relation from the spherical collapse model
to estimate radii of cores at this redshift. However, since cores are located in
elongated superclusters, this method most probably overestimates the mass. 
The increase of the mass of supercluster cores
can only come from the inflow of mass from the outer parts of a supercluster
\citep[see also][]{2014Natur.513...71T, 2019A&A...623A..97E}. 
We can find the mass in the outskirts of superclusters within turnaround
radius at $z = 0$ using observed mass distribution. 
Mass estimates of cores obtained in this way are in some cases lower 
than estimates based on simulations. This is because there is not 
enough mass in the outer parts of HDCs to increase mass twice
as predicted by simulations. 
We also find masses and radii of the regions in HDCs that may collapse in the future
and use them in comparison with local superclusters.
 
At redshift $z = 1.0$ simulations predict that
the masses of haloes have approximately five times
lower values than their present-day masses \citep{2015JKAS...48..213K}.
Using these mass values, we can approximately 
calculate the sizes of collapsing cores 
of superclusters at this redshift  using their present-day masses
(for local superclusters) and masses at redshift $z = 0.5$ for the 
BGW superclusters.

\section{HDCs of the BGW superclusters}
\label{sect:cores}

\subsection{Finding HDCs and their turnaround regions in the BGW superclusters}
\label{sect:collapse}

\citet{2017A&A...603A...5E} 
analysed  the luminosity density field in the BGW superclusters
and showed that the HDCs of superclusters can be 
separated at the density level
that  includes approximately $1/3$ of supercluster galaxies.
At this density level, the characteristic morphology of the superclusters
changes. We selected such regions in the BGW superclusters as candidates of the high-density
cores of superclusters. 
Then, we identified
high stellar mass galaxies in these regions with stellar masses of $\log(M_*/M_\odot)>11.3$
as the possible main galaxies of groups and clusters.
If there was more than one such galaxy, we chose 
the galaxy with the highest value of the stellar
mass as a possible centre of the HDC. 

We then drew spheres with an increasing
radii around these centre galaxies, and to obtain mass distribution around
these centres we calculated masses in the spheres using
the stellar masses of galaxies inside them with $\log(M_*/M_\odot)>11.3$. 
We also  corrected for the missing
galaxies, as described in Sect.~\ref{sect:masses}.
Next, we compared the distribution of mass in a core with the distribution of
mass obtained from the turnaround and future collapse 
mass-radius relations as described in Sect.~\ref{sect:sph} (Eqs.~\ref{eq:z05} and ~\ref{eq:mrfvs}).
We defined  regions  that 
have reached turnaround as regions with radii at which theoretical mass distribution
curve and mass distribution from observations crossed.
The future collapse regions were defined in the same way, using a mass-radius
relation for the future collapse.
To be selected as collapsing core of the supercluster,
the region has to 
contain at least two galaxies with stellar mass $\log(M_*/M_\odot)>11.3$
\citep[i.e. there should be 
at least two possible galaxy groups or clusters; see also][who defined massive
cores of superclusters as those which have at least two 
rich clusters]{2021A&A...649A..51E}.

Data concerning HDCs (their masses, sizes, and richness) are given in
Table~\ref{tab:tfc}, where masses and radii are given both for turnaround
and future collapse regions in HDCs. 
In this table, we also present estimated masses
of the HDC turnaround regions  at redshift $z = 0$.

\subsection{Masses and radii  of the collapsing regions in the HDCs}
\label{sect:collcores}

\begin{table*}%[ht]
\setlength{\tabcolsep}{6pt}
\caption{Data concerning HDCs in BGW superclusters.}
\center
\begin{tabular}{lrrrrrrrrrrrr} 
\hline\hline  
(1)&(2)&(3)&(4)&(5)& (6)&(7)&(8)&(9)&(10)&(11)&(12)&(13) \\      
\hline 
ID& $\mathrm{R.A.}$ & $\mathrm{Dec.}$  & $M_{\mathrm{T}}$ & $R_{\mathrm{T}}$ &
$D_{\mathrm{T}}$ &$N_{\mathrm{113}}$ &$N_{\mathrm{all}}$ & $M_{\mathrm{FC}}$ & $R_{\mathrm{FC}}$ 
& $M^{\mathrm{est}}_{\mathrm{mod}}$ & $R^{\mathrm{est}}_{\mathrm{mod}}$
& $M^{\mathrm{est}}_{\mathrm{obs}}$  \\
%&[deg]&[deg] &  [$10^{15}h^{-1}M_\odot$] & [$h^{-1}$ Mpc]  & 
%&  [$10^{15}h^{-1}M_\odot$] & [$h^{-1}$ Mpc]  &[$10^{15}h^{-1}M_\odot$] 
% & [$10^{15}h^{-1}M_\odot$] & [$h^{-1}$ Mpc] & [$10^{15}h^{-1}M_\odot$] \\
\hline
\multicolumn{2}{l}{BGW A} &&&&&&& \\
$1$& 163.51 & 55.37    & 0.96 & 4.5& 16.8 & 3 & 21 & 0.96 & 6.5 & 1.9 &  7.8 & 1.3  \\ 
$2$& 161.46 & 55.18    & 0.35 & 4.0& 15.2 & 2 & 8  & 0.42 & 4.5 & 0.8 &  5.9 & 0.7   \\ 
$3$&160.07  & 54.63    & 1.8  & 6.0& 10.0 & 2 & 22 & 1.8  & 7.3 & 3.6 &  9.6 & 1.8    \\ 
$4$& 154.47 & 52.38    & 1.6  & 5.8& 16.0 & 3 & 8  & 1.9  & 7.6 & 3.2 &  9.2 & 1.9    \\ 
\multicolumn{2}{l}{BGW B} &&&&&&& \\                               
$1$ & 163.10 & 50.14   & 0.53 & 3.5&  8.8 & 2 & 6  & 0.9  & 6.5 & 1.1 &  6.4 & 0.9      \\ 
$2$ & 163.00 & 47.32   & 0.81 & 4.5& 11.0 & 2 & 10 & 1.6  & 7.2 & 1.6 &  7.3 & 1.5      \\ 
\multicolumn{2}{l}{BGW C} &&&&&&& \\                                   
$1$& 168.81 & 53.33    & 3.3 &7.0&   15.0 & 2   & 17 & 3.3  & 9.5 & 6.6 & 11.7 & 3.3  \\ 
\multicolumn{2}{l}{BGW D} &&&&&&& \\                                   
$1$& 159.73 & 51.79    & 0.42 & 3.5& 11.3 & 2 & 11 & 1.0  & 6.4 & 0.8 &  6.4 & 1.0    \\ 
\hline
\label{tab:tfc}  
\end{tabular}\\
\tablefoot{                                                                                 
Columns are as follows:
(1) number of the region; 
(2)--(3) right ascension and declination of the region centre, in degrees;
(4) mass of the turnaround region at redshift $z = 0.5$, $M_{\mathrm{T}}$;
(5) radius of the turnaround region at redshift $z = 0.5$, $R_{\mathrm{T}}$;
(6) threshold luminosity density of the turnaround  region $D_{\mathrm{T}}$, 
in units of the mean density as described in the text;
(7) the number of galaxies with stellar masses $\log(M_*/M_\odot) \geq 11.3$
in the turnaround region;
(8) the number of all galaxies in the turnaround region;
(9)--(10) mass and radius of the future collapse region 
at redshift $z = 0.5$;
(11)--(12) estimated mass and radius of the turnaround region at redshift $z = 0$ 
(see text);
(13) mass within the region with size (column 12) estimated from observational data.
All masses are in units of $10^{15}h^{-1}M_\odot$ and radii in units of $h^{-1}$ Mpc.
}
\end{table*}

We determined four  HDCs 
in the BGW A supercluster, two in the BGW B supercluster, and one in each of the BGW C and D superclusters.
In Fig.~\ref{fig:radeccen}, we
plot the sky distribution of galaxies in the BGW superclusters and
show galaxies in HDCs  with different colours. The luminosity density limit for
HDCs in each supercluster, $D8_{HDC}$, is given in the caption. 
Figure~\ref{fig:radeccen}  shows the BGW superclusters as elongated systems, 
the BGW A and the BGW B superclusters being more elongated than BGW C and BGW D
\citep[details about the morphology of the BGW superclusters can be found 
in][]{2017A&A...603A...5E}.
Data about the HDCs of the BGW superclusters are given 
in Table~\ref{tab:tfc}.
Next, we describe HDCs in each BGW supercluster. 

\begin{figure}%[ht]
\centering
\resizebox{0.445\textwidth}{!}{\includegraphics[angle=0]{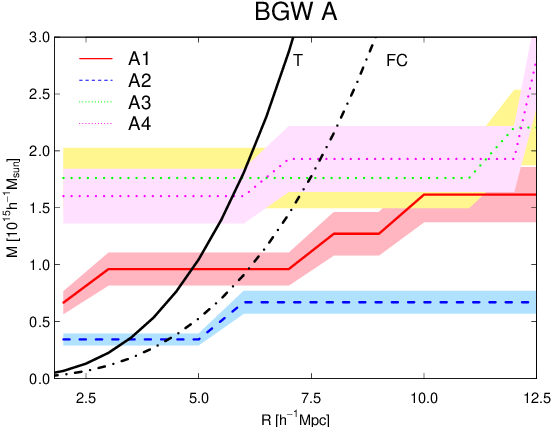}}
\caption{
Mass-radius relation for HDCs of BGW supercluster A.
Red solid line - A1; blue dashed line - A2; green dashed line - A3; violet 
dotted line - A4.
Black solid and dotted lines show turnaround (T) and future collapse (FC) mass 
($M_{\mathrm{T}}(R)$ and $M_{\mathrm{FC}}(R)$) 
versus radius of a sphere $R$ at redshift $z = 0.5$,
correspondingly.
Dashed areas show 15\% mass errors, as explained in the text. 
}
\label{fig:mra}
\end{figure}

The BGW~A supercluster has four HDCs. 
Figure~\ref{fig:mra} shows the distribution of masses in them.
Data on HDCs in BGW A are given in Table~\ref{tab:tfc}.
Core A1 is centred at the Planck cluster PSZ2 G151.62+54.78 with
the mass  $M \approx 5.4\times~10^{14}h^{-1}M_\odot$ \citep{2016A&A...594A..27P}. 
Approximately half of the mass here comes from this cluster.  
Table~\ref{tab:tfc} shows that the masses  in the turnaround regions of HDCs lie 
in the range of $M = 0.4 - 1.8\times~10^{15}h^{-1}M_\odot$, and their radii
are in the range of $R = 4 - 6$~\Mpc.
The masses and radii in regions that correspond
to the future collapse (calculated for the redshift $z = 0.5$) lie 
in the range of $M = 0.4 - 1.9\times~10^{15}h^{-1}M_\odot$, and 
$R = 4.5 - 7.6$~\Mpc.

The comparison with the total mass of the BGW A supercluster, 
given in Table~\ref{tab:bgwscl},
shows that turnaround regions of all four HDCs together 
contain  approximately 22\% of the total mass of the BGW A supercluster.
HDCs themselves contain approximately $1/3$ of the supercluster mass
(Sect.~\ref{sect:collapse}), which means that in some HDCs only the highest density 
central parts of the cores may collapse during the evolution of the supercluster. 
The prediction based on simulations (two-times increase of a halo mass)
probably overestimates  the
possible increase of core masses based on mass distribution
around haloes at redshift $z = 0$ 
in Table~\ref{tab:tfc}.  The mass distributions in Fig.~\ref{fig:mra} show that there are
no high-mass clusters at the outskirts of supercluster BGW A, which could join 
the cores and strongly increase their mass. This agrees with what
we know about local superclusters, where rich clusters are preferentially
located in the HDCs of superclusters, not in their outskirts
\citep{2005A&A...444..387D, 2005AdSpR..36..630B,
2020A&A...641A.172E, 2021A&A...649A..51E}.
This also suggests that mass estimates for the future collapse that use 
mass distribution around the centres of HDCs also present a better approximation
for the mass at redshift $z = 0$.
If we compare different mass estimates for the collapsing regions in the
HDCs of BGW A in Table~\ref{tab:tfc}, then at redshift $z = 0$
 cores 1 and 2 may have higher masses than predicted by the future collapse.

\begin{figure}%[ht]
\centering
\resizebox{0.445\textwidth}{!}{\includegraphics[angle=0]{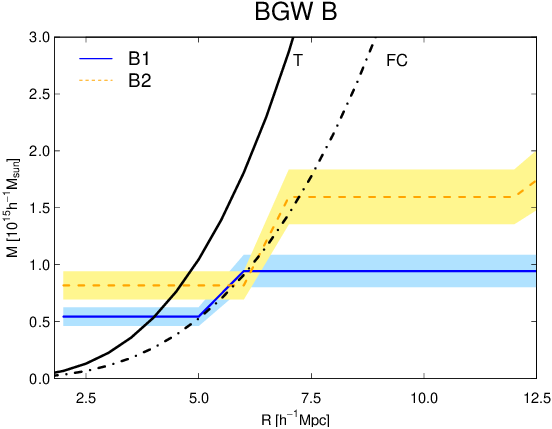}}
\caption{
Mass-radius relation for HDCs of BGW supercluster B.
Blue solid line - B1; red dashed line - B2.
}
\label{fig:mrb}
\end{figure}

\begin{figure}%[ht]
\centering
\resizebox{0.445\textwidth}{!}{\includegraphics[angle=0]{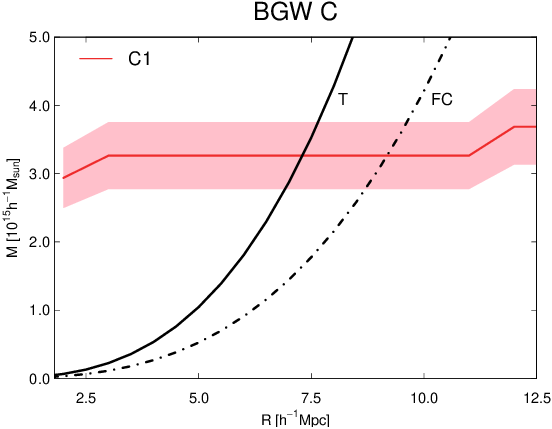}}
\caption{
Mass-radius relation for HDC of BGW supercluster C.
Red line - C1.
}
\label{fig:mrc}
\end{figure}

\begin{figure}%[ht]
\centering
\resizebox{0.445\textwidth}{!}{\includegraphics[angle=0]{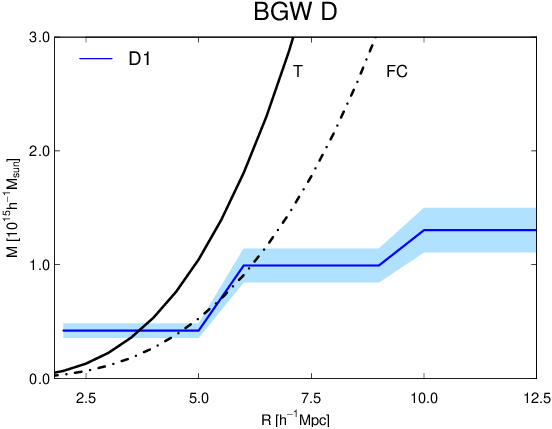}}
\caption{
Mass-radius relation for HDC of BGW supercluster D.
Blue line - D1.
}
\label{fig:mrd}
\end{figure}

In the BGW B supercluster, we determined two HDCs. Figure~\ref{fig:mrb}
shows the distribution of masses around them. From this figure, we can estimate that
the masses of their turnaround regions are $M = 0.5\times~10^{15}h^{-1}M_\odot$ and 
$0.8\times~10^{15}h^{-1}M_\odot$, and their radii
are $R = 3.5 $~\Mpc\ and $R = 4.5$~\Mpc.
In the future, their masses may be twice as high, 
$M = 0.9\times~10^{15}h^{-1}M_\odot$ and 
$1.6\times~10^{15}h^{-1}M_\odot$, and their radii
$R = 6.5$~\Mpc\ and $7.2$~\Mpc. These values are close to 
predictions based on observed mass distribution in Table~\ref{tab:tfc}. 
However, Fig.~\ref{fig:mrb} shows  that the increase of the mass for the future collapse
regions may be smaller, and their masses may stay in the range of 
$M \approx 0.5 - 0.9\times~10^{15}h^{-1}M_\odot$ 
and radii in the range of $R \approx 5 - 6$~\Mpc.

The total mass of the BGW B supercluster is $11\times~10^{15}h^{-1}M_\odot$
(Table~\ref{tab:bgwscl}), therefore 
the HDCs of the BGW B 
contain  approximately 12\% of the total mass of the supercluster at the turnaround.
During the future evolution the mass of cores may increase up to
approximately $1/4$ of the total mass of the supercluster. 
Thus, the structure and mass evolution within the BGW B is different from that 
of BGW A. 

We note that in Fig.~\ref{fig:radeccen} there is a dense clump of galaxies between
HDCs. This clump does not contain any high-mass galaxy with stellar mass 
$\log(M_*/M_\odot) \geq 11.3$ and thus it is not in the list of HDCs.
This clump is similar to, for example, high-density regions 
in the tail of the local supercluster SCl~A2142, which contain several pairs of
merging groups.
These are collapsing now and may become rich clusters or small superclusters
in the future.
These structures were described in detail in \citet{2018A&A...620A.149E} and
 \citet{2020A&A...641A.172E}.

The highest mass halo in supercluster C can be identified with the
Planck cluster PSZ2 G150.56+58.32  with the mass
of $M \approx 7.6\times~10^{14}h^{-1}M_\odot$ \citep{2016A&A...594A..27P}.
Among Planck clusters in the BGW redshift range, $0.47 \pm 0.05$,
this cluster has the highest estimated mass (Fig.~\ref{fig:mrc}). 
This is also the most massive cluster in the BGW superclusters.
The turnaround radius and core mass of the supercluster BGW C are 
the highest among
the HDCs of the BGW superclusters. The mass at the turnaround
is $M \approx 3.3\times~10^{15}h^{-1}M_\odot$,
which is approximately half of the total mass of the supercluster.
The radius of the core of the BGW C is the largest, $R \approx 7.0$~\Mpc. 
Figure~\ref{fig:mrc} shows that in a wide range of radii from the core centre,
the mass of the core does not increase, and the mass of the future collapse
region is the same as the mass of the turnaround region.
The mass in the future collapse region may actually be higher than our estimate,
since we cannot directly 
take into account possible poor groups in the supercluster.
However, the size of BGW C is over $60$~\Mpc, 
and we may assume that  its outskirts may separate from the collapsing core region in the future.

In BGW D (Fig.~\ref{fig:mrd}), the mass and size of the HDC 
at the turnaround
are similar to those of core B2 in BGW B, with a mass of 
$M \approx 0.4\times~10^{15}h^{-1}M_\odot$, and 
a radius of $R \approx 3.5$~\Mpc. Other rich clusters in BGW D, seen in
Fig.~\ref{fig:mrd}, may join its highest density core during the evolution, and
then its mass will be as high as $M \approx 1.0\times~10^{15}h^{-1}M_\odot$, and 
its radius $R \approx 6.4$~\Mpc.
The mass of the core at the turnaround forms approximately 13\% of the total
supercluster mass and may increase up to approximately $1/3$ of the supercluster
mass at the future collapse. 

In summary, we found that the masses of turnaround regions of the BGW superclusters
are in the range of $M_{T} = 0.4 - 3.3\times~10^{15}h^{-1}M_\odot$. The radii
of turnaround regions lie in the range of $R_{T} = 3.5 - 7$~\Mpc\ (Table~\ref{tab:tfc}).
The masses and sizes of their future collapse regions 
lie in the same range; however, for three HDCs the masses of the
future collapse regions are twice as high as in the turnaround region.
Thus, the increase of mass depends on the distribution of structures (galaxy groups
and filaments) in the HDCs.
At redshift $z = 1,$ their 
masses could have been $M_{T} \approx 0.2 - 1.6\times~10^{15}h^{-1}M_\odot$,
and their sizes are $R_{T} \approx 2.4 - 4.6$~\Mpc\ (Fig.~\ref{fig:mrz}).

\subsection{May HDCs merge in the future?}
\label{sect:merge}

\begin{figure}%[ht]
\centering
\resizebox{0.445\textwidth}{!}{\includegraphics[angle=0]{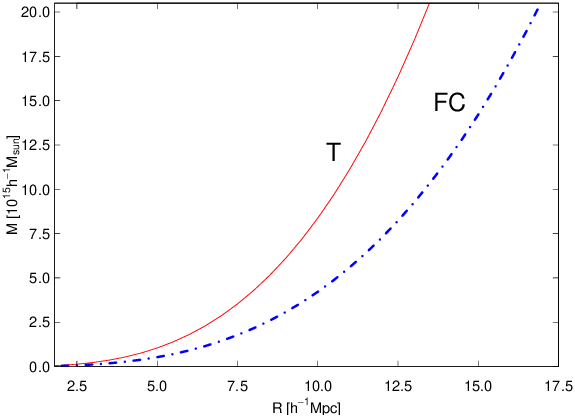}}
\caption{
Mass versus radius for the turnaround (T, red solid line) and future collapse
(FC, blue dot-dashed line) at redshift $z = 0.5$. 
}
\label{fig:mftfc15}
\end{figure}

We plot the sky distribution of galaxies from the
BGW superclusters in Fig.~\ref{fig:radeccen}, where  we also 
show the turnaround and future collapse regions of their HDCs. This figure 
summarises the results of the analysis of the HDCs.
FC regions may form separate systems in the future. 
Before making this conclusion, we intend to check whether different HDCs in the BGW A and B superclusters 
may still merge in the future. If this is a likely scenario, then they may 
eventually  form very massive and large systems similar to what 
\citet{2021A&A...649A..51E} found for the Corona Borealis supercluster
in the local Universe.
The answer to this question depends on the quality of our mass estimates of the  HDCs, 
which determine the radii of the turnaround and
future collapse regions, and on the mutual distances between HDCs.
If HDC masses are high enough to have overlapping 
core regions at the  turnaround, or which may collapse in the future,
then they may merge. 
In Fig.~\ref{fig:mftfc15}, we show the mass-radius relations for the turnaround
and future collapse epochs at redshift $z = 0.5$
(Eqs.~\ref{eq:z05} and ~\ref{eq:mrfvs}) for radii up to $17.5$~\Mpc. From this  we can find
how large the mass of a core should be in order to be at the  turnaround or to collapse in the future,
provided we know the radius of a region of interest.

Distances between individual cores in the BGW A (Fig.~\ref{fig:mra})
exceed $25$~\Mpc. The sizes of neighbouring future collapse regions
in BGW A are much smaller (Table~\ref{tab:tfc}). 
In order to overlap, the sizes of individual future collapse regions
in HDCs should be larger than $ \approx 13$~\Mpc. From Fig.~\ref{fig:mftfc15},
one can see that the mass needed for the future collapse in regions with the 
radius of $12 - 15$~\Mpc\ 
is at least of the order of $10 - 15\times~10^{15}h^{-1}M_\odot$. 
This is of the same order as the mass of the full BGW A superclusters,
$21 - 23\times~10^{15}h^{-1}M_\odot$ \citep{2017A&A...603A...5E}. 
For the BGW A 
supercluster, this means that to be that large, even the mass of the most massive core
($1.8\times~10^{15}h^{-1}M_\odot$) must be underestimated by almost ten times its actual value.

Distances between
cores of BGW B supercluster  (Fig.~\ref{fig:mrb}) exceed $30$~\Mpc.
In order to overlap, each of the HDCs should have  masses higher than 
$15\times~10^{15}h^{-1}M_\odot$ (Fig.~\ref{fig:mftfc15}).
This is higher than the total mass of the BGW B supercluster (Table~\ref{tab:bgwscl}),
$11 - 18\times~10^{15}h^{-1}M_\odot$, 
and six times higher than our estimate for the sum of the masses of the future
collapse regions of BGW B.

The mass estimates  of the BGW superclusters are, of course, more uncertain
than the mass estimates of the local superclusters. However, 
\citet{2017A&A...603A...5E} determined masses of the BGW superclusters
applying several methods, which included 
a stellar mass-halo mass relation for the first-rank galaxies, as used in this paper,
as well as mass-to-light ratios and combined morphological and physical
parameters of superclusters to determine supercluster masses. 
Different mass estimates were in a good agreement (see also Sect.~\ref{sect:masses}).
Also, mass estimates of the BGW superclusters in \citet{2017A&A...603A...5E} gave the total mass
of the whole BGW $\approx 40-50\times~10^{15} M_\odot$. This is slightly higher than the
mass of the most massive superclusters in, for example, 
\citet{2014A&A...567A.144C}, which is expected
since the BGW is a complex of four rich superclusters. If the mass of the BGW were
ten times higher, then its mass-to-light ratio should also be ten times higher than the
value found in \citet{2017A&A...603A...5E}, $M/L \approx 300$ (see Table~\ref{tab:bgwscl}). 
It is highly unlikely to have $M/L$ for superclusters ten times higher than that.
Our mass estimates for the HDCs with Planck clusters agree well with the masses
of these clusters (approximately half of the HDC mass comes from the mass of Planck clusters).
Therefore, we may assume that the HDCs in BGW A and BGW B 
never join, and during the evolution BGW A and BGW B 
will form several separate systems.
Of course, we need future studies
of superclusters at redshifts around $z = 0.5$ to better understand their properties.

\section{Comparison with local superclusters}
\label{sect:loc} 

\begin{figure}%[ht]
\centering
\resizebox{0.445\textwidth}{!}{\includegraphics[angle=0]{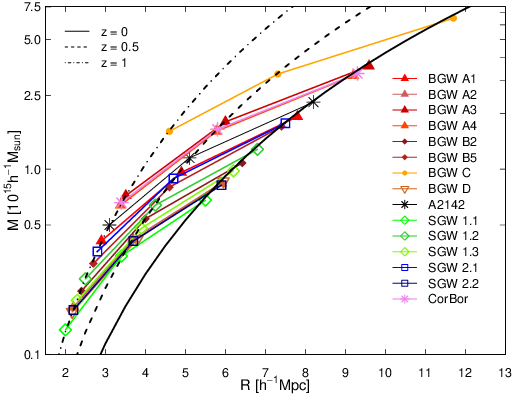}}
\caption{
Turnaround mass versus the radius of the turnaround region in the HDCs of superclusters
at different redshifts. For each HDC, the turnaround mass and radius 
are plotted at redshifts $z = 0$ (right point), $z = 0.5$ (middle point), and
$z = 1$ (left point). 
Legend lists HDCs of superclusters plotted in the figure.
Black lines show theoretical mass - radius relation for redshifts
$z = 0.0$ (solid line), $z = 0.5$ (dashed line), and $z = 1.0$ (dot-dashed line)
(Sect.~\ref{sect:sph}). 
}
\label{fig:mrz}
\end{figure}

Next, we compare the masses and radii of BGW supercluster HDCs with those 
of local superclusters (Fig.~\ref{fig:mrz}). For this comparison, we used  
data concerning local superclusters from the supercluster and group 
catalogues by \citet{2012A&A...539A..80L} and  \citet{2014A&A...566A...1T}. 
These catalogues are based on the MAIN sample of the 8th and 10th data 
releases of the SDSS (\citet{2011ApJS..193...29A} and \citet{2014ApJS..211...17A}. 
To compile group and supercluster
catalogues, an apparent Galactic extinction-corrected $r$ magnitude limit $r \leq 
17.77$ and redshift limits $0.009 \leq z \leq 0.200$ were applied. 
The redshifts of galaxies were corrected with respect to our motion relative to the
CMB and the comoving distances  of galaxies were computed, as described in
\citet{2002sgd..book.....M} and in \citet{2014A&A...566A...1T}.

Galaxy superclusters were determined as the BGW superclusters using the luminosity
density field of galaxies.   
The calculation of the luminosity density field and determination of
superclusters is described in detail in   \citet{2012A&A...539A..80L}.

To compare the masses and radii of the HDCs in the BGW (at redshift $z = 0.5$) 
and local superclusters (at redshift $z = 0$) in Fig.~\ref{fig:mrz},
we used data regarding the two richest superclusters in the SGW
\citep{2016A&A...595A..70E}, the Corona Borealis (CorBor) supercluster
\citep{2021A&A...649A..51E}, and the supercluster SCl~A2142
\citep{2018A&A...620A.149E}.
The SGW is a complex of two rich and three poor superclusters.
We used data regarding three HDCs in the richest SGW supercluster (denoted as 
SGW 27) and two cores in the second richest SGW supercluster (SGW 19).
In all these studies, the masses and radii of HDCs
of superclusters correspond to the masses and radii of the turnaround regions
of supercluster cores, they were found using a similar methodology,
and therefore we can directly compare their estimates. 
To compare HDCs at different redshifts, one needs to convert corresponding 
masses and radii to the same redshift. 
This is done as described in Sect \ref{sect:masses}.

In Fig.~\ref{fig:mrz}, we plot the turnaround mass versus radius for the 
the cores of superclusters at different redshifts. 
For each supercluster, the turnaround mass and radius 
are plotted at redshifts $z = 0$ (right point), $z = 0.5$ (middle point), and
$z = 1$ (left point). 
For the BGW supercluster cores, estimated masses and radii at redshift $z = 0$
are given using predictions from simulations (Table~\ref{tab:tfc}, columns 11 and 12).
Black lines in the figure show a theoretical mass - radius relation for redshifts
$z = 0.0$, $z = 0.5$, and $z = 1.0$. 
This figure  shows the growth of the supercluster cores that have reached
the  turnaround and started to collapse.

In Fig.~\ref{fig:mrz}, the turnaround region of the HDC of the 
BGW C supercluster has the highest mass and 
largest radius at all redshifts. 
Other turnaround regions of the HDCs of the BGW superclusters are of the same order as
turnaround regions in the Corona Borealis and A2142 superclusters. The collapsing cores
in the SGW superclusters are less massive.

\begin{figure}%[ht]
\centering
\resizebox{0.445\textwidth}{!}{\includegraphics[angle=0]{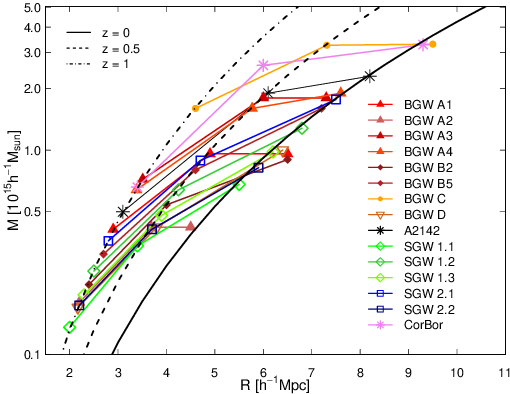}}
\caption{
Future collapse mass versus the radius of the HDCs of the BGW superclusters
at redshifts $z = 0$ (right point; 
columns 9 and 10 in Table~\ref{tab:tfc}), $z = 0.5$ (middle point), and
$z = 1$ (left point), and the masses and radii of the core of the Corona Borealis supercluster
and SCl~A2142, which at $z = 0.5$ correspond to those of their spheres of influence
(see text).
Legend lists HDCs of superclusters plotted in the figure.
Black lines show theoretical mass - radius relation for redshifts
$z = 0.0$ (solid line), $z = 0.5$ (dashed line), and $z = 1.0$ (dot-dashed line)
(Sect.~\ref{sect:sph}). 
}
\label{fig:mrfcz}
\end{figure}

However, the two-fold increase of the HDC masses
of the BGW superclusters in Fig.~\ref{fig:mrz}, 
as predicted in simulations, does not take into account the mass distribution in the HDCs and 
is probably overestimated for five cores (for BGW A cores, and for BGW C).
Also, this prediction is valid for individual haloes, not for HDCs of superclusters
analysed in this paper.
We argued that the masses and radii of the future collapse regions may also be better indicators for 
masses and radii of collapsing regions in the BGW superclusters at redshift $z = 0$. 

For the same reason, the decrease of 
masses and radii of the cores of local superclusters at high redshifts may be 
overestimated. We can use another estimate based on the spheres of influence around 
rich clusters in superclusters.
\citet{2020A&A...641A.172E} and \citet{2021A&A...649A..51E}
showed that rich clusters in the HDCs of supercluster 
SCl~A2142 and the Corona Borealis are surrounded
by a sphere of influence, where all galaxies and groups of galaxies are 
infalling towards clusters. The masses and sizes of the spheres of influence are of the same
order as the masses and sizes of the turnaround regions of the HDCs
of the BGW superclusters. The density contrast at the borders
of the spheres of influence is $\Delta\rho \approx 30$. 
\citet{2020A&A...641A.172E} and \citet{2021A&A...649A..51E} found that
the spheres of influence passed the  turnaround and started to collapse 
at redshifts $z \approx 0.4$,  approximately at the redshift
of the BGW. We may assume that the HDCs
of the BGW superclusters may be considered the progenitors of the 
spheres of influence around clusters in the HDCs of local superclusters.
Therefore,  in Fig.~\ref{fig:mrfcz} we compare the masses and radii of the 
future collapse regions of the HDCs of the BGW 
superclusters (Table~\ref{tab:tfc}) and those of the spheres of influence (for SCl~A2142
and for the Corona Borealis supercluster; for the SGW cores, we used the same values as in
Fig.~\ref{fig:mrz}).

As Fig.~\ref{fig:mrfcz} shows, the differences between the masses and sizes of 
the BGW superclusters and of other superclusters calculated in this way are smaller
than those seen in Fig.~\ref{fig:mrz}.
Thus, it is possible that the BGW superclusters will evolve to systems
similar to the (HDCs of the) local rich superclusters.
For example, even BGW C, which has the most massive core among the
cores of the BGW superclusters, will probably have a mass close to that of the
HDC of the Corona Borealis supercluster 
\citep[for details, see][]{2021A&A...649A..51E}.
The HDCs of other BGW superclusters will evolve to systems similar in mass
and size to the HDCs of SCL~A2142 and of the SGW superclusters.

\section{Discussion and summary}
\label{sect:disc} 

\subsection{Comparison with other superclusters at redshifts of $z = 0.5$ and $z = 1$}
\label{sect:hiz} 

Are the BGW superclusters typical or exceptional among superclusters at higher redshifts?
To answer this question, we briefly  compare the BGW superclusters with other superclusters  
at redshifts of approximately $z = 0.5$ and $z = 1$. 

\citet{2011A&A...532A..57S} analysed the mass and structure
of supercluster SCl2243 at redshift $z = 0.447$. 
They found that the mass of the central cluster of the supercluster
is of the order of $M \approx 1.5\times10^{15}M_\odot$. 
This is higher than the estimated mass of individual clusters  in the BGW
(including Planck clusters). 
The mass of filaments around the central cluster was found to be of the same order.
Together, the total mass of SCl2243 is $M \approx 3\times10^{15}M_\odot$,
which is of the same order as the total mass of the BGW D supercluster,
or as the HDC of the BGW C supercluster.
In SCl2243, clusters are connected by
several galaxy filaments that are similar in morphology to the
BGW C supercluster, described in \citet{2017A&A...603A...5E}.

\citet{2012MNRAS.421.1949V} analysed the galaxy and group content
and large-scale structure around a very rich galaxy cluster  
at redshift $z = 0.45$. They found that
large-scale structures around the cluster extend up to approximately 
$20$~Mpc. In this complex, the fraction of blue galaxies decreases with the local
density. The galaxy content of groups shows a large variation.
They concluded that the large-scale structure around a rich cluster 
is a dynamical place for galaxy evolution, as also
found in the studies  of other superclusters \citep{2018A&A...620A.149E}.

\citet{2018MNRAS.475.4148G} discovered a galaxy supercluster at redshift $z = 0.65$
and confirmed with VLT/VIMOS spectroscopic observations that it
has four galaxy clusters and approximately ten groups.
The masses of clusters are of the order of $M \approx 10^{14}M_\odot$,
which is slightly lower than the masses of  Planck clusters
in the BGW superclusters. The mass growth of groups in
this supercluster  from redshift $z = 0.65$ to $z = 0.46$
\citep[approximately 1.5 times; see][]{2015JKAS...48..213K}
may lead to systems with the same order of mass
as the turnaround regions of the HDCs of the BGW superclusters. 

\citet{2007MNRAS.379.1546T} found a rich filament of galaxies 
at redshift $z \approx 0.55$ connecting clumps of galaxies
with a massive cluster (with  $M \approx 10^{15}M_\odot$).
These clumps are probably galaxy groups aligned along
the filament and probably bound to the main cluster.  

\citet{2017ApJ...844...25B} reported the discovery of a very rich supercluster
at redshift $z \approx 0.3$, which is nicknamed Saraswati. They estimated that the core region
within turnaround radius 
of the Saraswati with very rich Abell cluster \object{A2631} has a mass of 
$M \approx 4\times10^{15}M_\odot$ and comoving radius of $R \approx 20$~Mpc.
The whole Saraswati supercluster is more massive and larger than the Corona Borealis supercluster,
which is the most massive among local superclusters used for comparison in this study.
Based on the predictions from simulations, we can estimate that the mass of the turnaround region
of the Saraswati supercluster at redshift $z \approx 0.5$ could be 
$M \approx 3\times10^{15}M_\odot$, which is similar to the turnaround region in the 
BGW C supercluster. There are other similarities between BGW C and Saraswati.
There is  one very massive galaxy cluster in the  cores of both superclusters. The 
masses of clusters are
of the same order, $M \approx 10^{15}M_\odot$ 
\citep{2021MNRAS.501..756M}. Both of these clusters are also 
Planck clusters: PSZ2 G150.56+58.32 in BGW C
and PSZ2 G087.03-57.37 (A2631) in Saraswati. 
It could be an interesting future task to compare Saraswati and BGW C in more detail.

\begin{figure*}[ht]
\centering
\resizebox{0.44\textwidth}{!}{\includegraphics[angle=0]{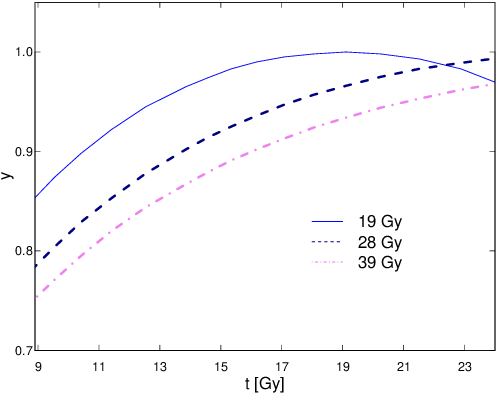}}
\resizebox{0.44\textwidth}{!}{\includegraphics[angle=0]{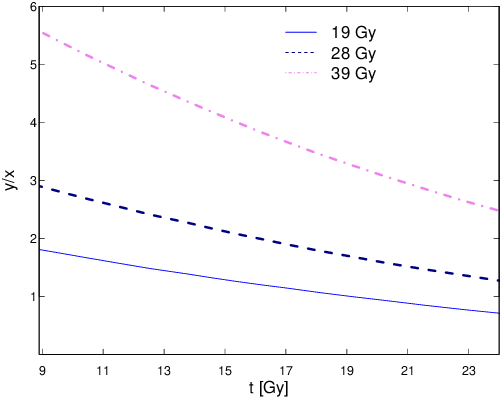}}
\caption{
Evolution of radius $R$ of future collapse region with the age of the universe.
In the left panel, $y$ denotes the ratio $y = R/R_{T}$, 
where $R$ is the radius of a overdensity region 
and $R_{T}$  is its radius at a turnaround time.
In the right panel, $x$ denotes the ratio of the usual cosmological scale factor, $a$,
to the turnaround scale factor, $a_{T}$ ($x = a/a_{T}$).
}
\label{fig:rfc}
\end{figure*}

The masses of superclusters at redshift $z \approx 0.5 $ from these studies are
of about $0.1 - 5\times~10^{15} M_\odot$. A comparison 
with the masses of HDCs of the BGW superclusters
and predicted masses of local superclusters (Table~\ref{tab:tfc} and Fig.~\ref{fig:mrz})
shows that these masses are of the same order, and we may suppose
that superclusters, currently observed at $z \approx 0.5$,
will evolve into systems that are of the same size and mass as local
rich superclusters or their HDCs.

\citet{2016ApJ...821L..10K} reported a discovery of a supercluster 
at redshift $z \approx 0.9$ that consists of three massive galaxy clusters
with separations of approximately $ 15$~Mpc and masses of the order of
$0.1 - 5\times~10^{15} M_\odot$. This is the same mass range 
as the estimated masses of superclusters in our study.

\citet{2008ApJ...677L..89G} and \citet{2013ApJ...768..104F} 
analysed a compact supercluster of  three
massive clusters at redshift $z \approx 0.9$ (RCS 2319+00), with X-ray masses
of the order of $\approx 5\times~10^{14} M_\odot$. The authors proposed that these clusters may merge
and form a massive cluster at lower redshifts. The mass of clusters is
close to our estimate for the cores of the BGW superclusters, and 
these clusters may represent a structure similar to the HDCs of 
superclusters.

One of the largest superclusters discovered at redshift $z \approx 0.9$ 
is the supercluster CL1604 \citep{2019PASJ...71..112H}.
This supercluster has three clusters with masses of $M > 10^{14} M_\odot$ and 
several less massive groups \citep{2019PASJ...71..112H}. The size of this structure
is $\approx 26$ comoving Mpc, embedded in a larger supercluster. Thus, clusters in the CL1604
supercluster may form a core region that is part of the larger structure with several galaxy
groups.

\subsection{Evolution of superclusters in the cosmic web}
\label{sect:evol}

The studies of protoclusters suggest that the changes in the
masses and radii of regions around protoclusters from where they collect their
matter are approximately in the same range as the 
changes of supercluster core masses and radii in 
Fig.~\ref{fig:mrfcz} \citep{2013ApJ...779..127C}.
\citet{2013ApJ...779..127C} analysed the assembly of protoclusters
and calculated the growth of their mass and the size of regions from
where clusters obtained their mass via the inflow of groups and galaxies.
At redshift $z = 1,$ the richest clusters with present-day masses of the order of
$M \approx 10^{15}h^{-1}M_\odot$ collect their mass from regions with
sizes of $R  \approx 4$ comoving Mpc, which is in good agreement with the prediction of the sizes
of the HDCs of superclusters at $z = 1$ found in this study. 

In Fig.~\ref{fig:rfc}, we show the size evolution  of the radius $R$ of 
three different future collapse regions with turnaround times 
at $19$, $28$, and $39$ Gyr (from the Big Bang).
We used the dimensionless designations by \citet{1998ApJ...508..483W}: 
$x = a/a_{T}$, $y = R/R_{T}$. Here, $a$ is the usual cosmological scale factor, 
$R$ denotes the radius of an overdensity region, and
$a_{T}$ and $R_{T}$ are corresponding values at the time of the turnaround.

The left panel of Fig.~\ref{fig:rfc} shows how the radius of the overdensity
region  $R$ (in units of the radius at the turnaround)
changes with time.
For example, if the turnaround occurs at $19$~Gyr, then this 
ratio increases $\approx 1.14$ times from redshift $z = 0.5$ ($\approx 5$~Gyr ago)
to the present time, and continues to increase during next  $5$~Gyr.
However, the increase of radius slows down, and 
in total, during these $10$~Gyr this ratio increases $\approx 1.15$ times only.
In the right panel of Fig.~\ref{fig:rfc} vertical axis can be taken as the evolution of radius of 
the overdensity region relative to the overall cosmological expansion. 
This presentation shows  that from $z=0.5$ to the present day, the 
relative radius of the overdensity region has decreased by about $1.3$ times 
for all three cases. Over $10$~Gyr, it decreased by approximately $1.8$ times.

Similarly, 
\citet{2013ApJ...779..127C} showed that the effective sizes and masses 
of regions around protoclusters
almost do not change from redshift $z = 0.5$ to redshift $z = 0$.
The actual changes depend on the structures in cores around clusters,
as we show above. A large variety of the properties of superclusters cores 
is similar to what have been  
found for protoclusters from Millenium simulations 
where the final mass
of the cluster does not depend strongly on the mass of a most massive 
progenitor cluster at redshift 2 \citep{2015MNRAS.452.2528M}.
However, the estimations of the changes of the radii of the turnaround regions
show that they evolve slowly, 
and it is unlikely that such dense and rich superclusters at redshift
$z \approx 0.5$ could fall apart fast enough to form poorer superclusters at $z = 0$. 
Thus, the current tension remains.
In this case, the tension between the  morphology
of the BGW and local superclusters also remains.

Some HDCs of superclusters may merge and form a single cluster
in the future \citep{2009MNRAS.399...97A}. Simulations show that
approximately half of the mass of present-day massive clusters 
comes from haloes that merge with the protoclusters at redshifts $z < 1$
\citep{2015JKAS...48..213K, 2015MNRAS.452.2528M}. We can look at the HDCs 
of superclusters as future rich clusters that are presently still forming.
In the local superclusters, such HDCs with several rich clusters
that may merge in the future have been found, for example,
in the Corona Borealis supercluster \citep{2021A&A...649A..51E},
and in the Shapley supercluster \citep{2005A&A...444..387D}. The cores of the
BGW superclusters may represent such systems. They themselves are too far apart 
to merge, but each core may turn into a rich cluster in the distant future.

We may also ask what observational signatures of a turnaround region around the HDC
of a supercluster could be, considering that we to not have precise data on peculiar velocities of galaxies
from observations. For the supercluster SCl~A2142, \citet{2018A&A...620A.149E} and
\citet{2020A&A...641A.172E} found an excess of star-forming galaxies 
at its turnaround region. 
Moreover, \citet{2020A&A...641A.172E} showed that long galaxy filaments that
surround the main body of SCl~A2142 discontinue at the turnaround region, and
they are detached from the inner regions of the supercluster.
The regions of influence around clusters
for which the turnaround occurred at redshift $z \approx 0.4$ is seen
by the (rather weak) minimum in the galaxy distribution around them \citep{2020A&A...641A.172E,
2021A&A...649A..51E}. Thus, one needs to study the structure and galaxy distribution 
of superclusters in detail to understand their dynamical state.

\subsection{Summary}
\label{sect:sum} 

In this study, we analysed the luminosity density distribution in the 
BGW superclusters and identified their HDCs. 
We used the spherical collapse model to find the masses and sizes
of the turnaround regions of the HDCs of the BGW superclusters and to
predict their evolution. 
A comparison with the spherical collapse model suggested
that supercluster cores are already collapsing. 
Our study of the properties of these cores can be summarised as follows.

\begin{itemize}
\item[1)]
We determined eight  HDCs in the BGW superclusters.
The masses of their turnaround regions 
are in the range of $M_{\mathrm{T}} \approx 0.4 - 3.3\times~10^{15}h^{-1}M_\odot,$
and radii are in the range of  $R_{\mathrm{T}} \approx 3.5 - 7$~\Mpc. 
\item[2)]
The masses of the future collapse regions of five cores are in the same range 
as the turnaround masses, and for three cores 
the mass of the future collapse regions may increase twice. 
The radii of their future collapse regions
are somewhat larger,  in the range of  $R_{\mathrm{FC}} \approx 4 - 8$~\Mpc. 
\item[3)]
The  mass estimates obtained for the future collapse regions are also good
approximations for the possible masses of the collapsing regions
of the BGW HDCs at redshift $z = 0$.
\item[4)]The masses and sizes of the regions in the BGW superclusters 
that may eventually
collapse are comparable with those
of the massive collapsing supercluster cores in the local Universe.
\end{itemize}

The evolution of an elongated supercluster with several HDCs connected by
filamentary structures occurs under the influence of the  
gravitational pull of the cores and accelerated
expansion of the Universe. Our analysis, based on
the spherical collapse model, suggests that the HDCs of the BGW superclusters are collapsing
and pulling the intervening galaxies in their directions. 
Thus, the filamentary structures between the
cores may break down. This  produces a gap between HDCs. Therefore,
it is likely that separate superclusters  will form.
The masses and sizes
of these superclusters are comparable to those of local rich  and medium-rich
superclusters. {\bf This finding may weaken the tension with the
$\Lambda$CDM model, which does not predict a large number of rich superclusters
in our local cosmic neighbourhood. }
This also explains why there are no superclusters as elongated as 
the BGW superclusters in the local Universe - such superclusters are split 
into smaller, less elongated systems.
The mass and size of the collapsing core of the BGW C supercluster, which is 
the most massive HDC in the BGW,
will probably become similar to the Corona Borealis supercluster,
which is one of the largest and most massive
bound systems in the nearby Universe.

However, whether the tension with the $\Lambda$CDM model will weaken depends
on the timescale during which the HDCs collapse and form separate systems.
We showed that, for example, over $10$~Gyr ($5$~Gyr ago and $5$~Gyr into the future),
the radii of the turnaround regions do not change much. How large systems in the cosmic web
evolve can be studied with simulations. Also, 
future deep surveys such as J-PAS \citep{2014arXiv1403.5237B} 
will provide us with data concerning galaxy
clusters and superclusters in a wide range of redshifts up to at least
$z \approx 1$. This will make it possible to study the properties and
evolution of a large sample of galaxy clusters and superclusters and to better understand their evolution, the evolution of the whole cosmic web, and the 
properties of dark matter and dark energy.

\section*{Acknowledgments}

We thank the referee for comments and helpful suggestions.
The present study was supported by the ETAG projects 
PSG700, PRG1006,  PUT1627, PUTJD907,  and by the European Structural Funds
grant for the Centre of Excellence "The Dark Side of the Universe" (TK133).
SS acknowledges the support of the European Regional Development Fund and the
Mobilitas Pluss postdoctoral research grant MOBJD660.
This work has also been supported by
ICRAnet through a professorship for Jaan Einasto.

Funding for SDSS-III has been provided by the Alfred P. Sloan Foundation, 
the Participating Institutions, the National Science Foundation, 
and the U.S. Department of Energy Office of Science. 
The SDSS-III web site is http://www.sdss3.org/.
SDSS-III is managed by the Astrophysical Research Consortium 
for the Participating Institutions of the SDSS-III Collaboration 
including the University of Arizona, the Brazilian Participation Group, 
Brookhaven National Laboratory, Carnegie Mellon University, 
University of Florida, the French Participation Group, 
the German Participation Group, Harvard University, 
the Instituto de Astrofisica de Canarias, 
the Michigan State/Notre Dame/JINA Participation Group, 
Johns Hopkins University, Lawrence Berkeley National Laboratory, 
Max Planck Institute for Astrophysics, 
Max Planck Institute for Extraterrestrial Physics, 
New Mexico State University, New York University, Ohio State University, 
Pennsylvania State University, University of Portsmouth, 
Princeton University, the Spanish Participation Group, 
University of Tokyo, University of Utah, Vanderbilt University, 
University of Virginia, University of Washington, and Yale University. 
In this work we used R statistical environment
\citep{ig96}.

\bibliographystyle{aa}
\bibliography{bgwcollapse.bib}

\end{document}